\documentclass{article}
\usepackage{ais25}
\usepackage[T1]{fontenc}

\usepackage{url}
\usepackage{amsmath}
\usepackage{amssymb}
\usepackage{xcolor}

\title{The Cost-Benefit of Interdisciplinarity in AI for Mental Health}

\author{\name{Katerina Drakos\footnotemark[1]} \addr{Center for Social Data Science, Faculty of Social Sciences, University of Copenhagen} \email{kda@sodas.ku.dk} 
\\ \name{Eva Paraschou\footnotemark[1]} \addr{Dept. of Applied Mathematics and Computer Science, Technical University of Denmark} \email{evpa@dtu.dk}
\\ \name{Simay Toplu\footnotemark[1]} \addr{Institute for Ethics in Artificial Intelligence, School of Social Sciences and Technology, Technical University of Munich} \email{simay.toplu@tum.de}
\\ \name{Line Harder Clemmensen} \addr{Dept. of Mathematical Sciences, University of Copenhagen} \email{lkhc@math.ku.dk}
\\ \name{Christoph Lütge} \addr{Institute for Ethics in Artificial Intelligence, School of Social Sciences and Technology, Technical University of Munich} \email{luetge@tum.de}
\\ \name{Nicole Nadine Lønfeldt} \addr{Child and Adolescent Mental Health Center, Copenhagen University Hospital} \email{nicole.nadine.loenfeldt@regionh.dk}
\\ \name{Sneha Das} \addr{Dept. of Applied Mathematics and Computer Science, Technical University of Denmark} \email{sned@dtu.dk}
}

\begin{document}

\maketitle
\footnotetext[1]{These authors contributed equally to this work. Their names are listed in alphabetical order by surname.}

\begin{abstract}
Artificial intelligence has been introduced as a way to improve access to mental health support. However, most AI mental health chatbots rely on a limited range of disciplinary input, and fail to integrate expertise across the chatbot's lifecycle. This paper examines the cost-benefit trade-off of interdisciplinary collaboration in AI mental health chatbots. We argue that involving experts from technology, healthcare, ethics, and law across key lifecycle phases is essential to ensure value-alignment and compliance with the high-risk requirements of the AI Act. We also highlight practical recommendations and existing frameworks to help balance the challenges and benefits of interdisciplinarity in mental health chatbots.
\end{abstract}

\begin{keywords}
Interdisciplinary collaboration; mental health; chatbot; human-computer interaction
\end{keywords}

\section{Motivation \& Background}

Mental health care faces a global crisis, marked by rising demand (\cite{world2022highlights}), especially among young people following COVID-19 (\cite{xiong2020impact}), unequal access and increased stigma (\cite{wainberg2017challenges}). To increase accessibility, artificial intelligence (AI) has been applied in different mental health care services. Some applications focus on mental health literacy and offer psychoeducational content (\cite{potts2023multilingual}). Others stimulate therapeutic interaction to help users manage their mental health (\cite{de2025ai}; \cite{thieme2023designing}; \cite{fitzpatrick2017delivering}). These mental health chatbots can serve as low-threshold entry points to care, increasing accessibility particularly for underserved populations (\cite{pozzi2025keeping}) and individuals hesitant to seek professional help (\cite{hoffman2024understanding}). 

AI in mental health care must address usability and clinical benefit, yet no discipline can achieve this alone (\cite{boulos2021iterative}; \cite{werth2025success}). Developing mental health chatbots requires expertise from multiple disciplines, such as health, psychology, psychiatry, human-computer interaction (HCI), computer science, ethics, and law (\cite{kuhn2024interdisciplinary}; \cite{spahl2025integrating}; \cite{floridi2018ai4people}; \cite{morrin2025delusions}). Interdisciplinary teams help ensure contextually appropriate chatbots (\cite{pagliari2007design}), and generate new cross-sectoral research directions (\cite{molina2016role}). Collaboration is crucial, as chatbots built from a single disciplinary perspective can overlook risks such as privacy, safety and accountability concerns which demand joint expertise (\cite{floridi2018ai4people}). Thus, a multilevel, justice-oriented and interdisciplinary approach is essential to ensure inclusivity and responsiveness to societal needs (\cite{kuhn2024interdisciplinary}). 

A 2023 survey on conversational agents in mental health revealed that only 19.6\% of chatbots employed interdisciplinary teams (\cite{cho2023integrative}). Exploring the current cost-benefit trade-off of interdisciplinary collaboration in this field, Table~\ref{tab:chatbots} provides a short overview of how experts from technology, healthcare, ethics or law (as suggested in the literature (\cite{kuhn2024interdisciplinary}; \cite{spahl2025integrating};\cite{floridi2018ai4people}; \cite{morrin2025delusions})) are integrated across mental health chatbots' lifecycle (design, development, evaluation). We employed a convenience sampling strategy on papers from 2022-2025 (to build on the findings of \cite{cho2023integrative}) from MDPI with inclusion criteria consisting of "chatbot" and "mental health" in the title or keywords, as well as on other papers from different publishers presenting chatbots that explicitly applied interdisciplinary collaboration. After abstract screening of 16 papers, Table~\ref{tab:chatbots} presents the papers included in our overview.

\begin{table}[ht!]
\tiny
\caption{An overview of interdisciplinary collaboration in recent chatbots. The letters represent the first letter of each expert group.}
\label{tab:chatbots}
\begin{tabular}{c|cccc|ccc}
\hline
\multicolumn{1}{l|}{} & \multicolumn{4}{c|}{\textbf{Experts}} & \multicolumn{3}{c}{\textbf{Lifecycle Phase}} \\ \hline
\textbf{Chatbot} & \multicolumn{1}{c|}{Technology} & \multicolumn{1}{l|}{Healthcare} & \multicolumn{1}{l|}{Ethics} & \multicolumn{1}{l|}{Law} & \multicolumn{1}{c|}{Design} & \multicolumn{1}{c|}{Development} & Evaluation \\ \hline
\cite{manole2024harnessing}  & \multicolumn{1}{c|}{X}  & \multicolumn{1}{c|}{\checkmark} & \multicolumn{1}{c|}{X}  & X  & \multicolumn{1}{c|}{-}   & \multicolumn{1}{c|}{-}  & -  \\ \hline
\cite{kamdan2025early}  & \multicolumn{1}{c|}{\checkmark} & \multicolumn{1}{c|}{X}  & \multicolumn{1}{c|}{X} & X  & \multicolumn{1}{c|}{-} & \multicolumn{1}{c|}{-}  & - \\ \hline
\cite{rathnayaka2022mental}  & \multicolumn{1}{c|}{\checkmark} & \multicolumn{1}{c|}{\checkmark} & \multicolumn{1}{c|}{\checkmark} & X                 & \multicolumn{1}{c|}{T-H-E}  & \multicolumn{1}{c|}{-}  & -   \\ \hline
\cite{hall2022sustainable}  & \multicolumn{1}{c|}{\checkmark}   & \multicolumn{1}{c|}{\checkmark} & \multicolumn{1}{c|}{X}  & X   & \multicolumn{1}{c|}{T-H}  & \multicolumn{1}{c|}{T-H}  & T-H  \\ \hline
\cite{noble2022developing} & \multicolumn{1}{c|}{\checkmark}  & \multicolumn{1}{c|}{\checkmark}   & \multicolumn{1}{c|}{X}  & X & \multicolumn{1}{c|}{T-H}  & \multicolumn{1}{c|}{T-H}   & T-H   \\ \hline
\cite{chua2023parentbot} & \multicolumn{1}{c|}{\checkmark} & \multicolumn{1}{c|}{\checkmark} & \multicolumn{1}{c|}{X} & X & \multicolumn{1}{c|}{T-H}   & \multicolumn{1}{c|}{T-H}  & T-H  \\ \hline
\cite{olla2025deploying}  & \multicolumn{1}{c|}{\checkmark}  & \multicolumn{1}{c|}{\checkmark} & \multicolumn{1}{c|}{\checkmark} & X  & \multicolumn{1}{c|}{T-H-E}  & \multicolumn{1}{c|}{T-H-E}  & T-H-E \\ \hline
\end{tabular}
\end{table}

Despite some examples of integration, some chatbots still operate without interdisciplinary input, focusing solely on technology or healthcare (\cite{manole2024harnessing}; \cite{kamdan2025early}), potentially reflecting the challenges outlined earlier. Most reviewed chatbots (\cite{noble2022developing}; \cite{hall2022sustainable}; \cite{chua2023parentbot}) include only technology and healthcare experts across all lifecycle phases. While \cite{rathnayaka2022mental} integrates experts in technology, healthcare, and ethics, this interdisciplinary collaboration is limited to the design phase, whereas \cite{olla2025deploying} extend this collaboration throughout the lifecycle. {\it In sum, our mental health chatbots overview (Table \ref{tab:chatbots}) suggests a potential cost-benefit trade-off: even where interdisciplinary teams exist, collaboration often appears limited to specific phases and does not cover a wide range of disciplines.}

Implementing interdisciplinarity in mental health chatbots is often difficult. Challenges include misaligned expectations, limited cross-disciplinary training, conflicting timelines, and tensions between business and scientific standards (\cite{pagliari2007design}). Communication barriers such as different terminology and professional priorities add to these challenges (\cite{doherty2010design}; \cite{blandford2018seven}). Structural barriers such as restricted funding across fields (\cite{cibrian2022interdisciplinary}) and publishing norms discouraging interdisciplinary work (\cite{yegros2015does}) further constrain collaboration, reducing adoption and impact throughout the chatbot's lifecycle (\cite{cibrian2022interdisciplinary}).

\section{Position Statement}

Despite recognizing the challenges of interdisciplinary collaboration in mental health chatbots, {\it we argue for deliberately embedding technology, healthcare, ethics, and law experts in the most impactful lifecycle phases of mental health chatbots, to guide ethical and compliant implementation.} Our position is founded on the recent EU AI Act (\cite{EUR-Lex_52021PC0206}), which classifies mental health AI chatbots as high-risk and sets strict requirements for transparency, human oversight and risk management. Meeting these obligations requires expertise in technology, healthcare, and law to accurately translate regulatory standards into practice (\cite{montag2024successful}). Moreover, frameworks such as Value Sensitive Design (\cite{friedman1996value}) and ethics-by-design (\cite{van2020designing}), encourage the integration of ethical and social values throughout the systems' (including chatbots') lifecycle, underscoring additional input from ethicists. Therefore, we argue that meeting regulatory standards can be supported by these ethical frameworks by guiding phased contributions from legal and ethical experts alongside technical and clinical teams. 

In this position paper, we address the broad mental health domain instead of delving deeper into specific mental health disorders. Thus, our position emphasizes the generic disciplines highlighted as the most influential ones in the literature of digital health interventions for mental health. We acknowledge that each individual mental health condition requires additional expertise in the specific field. An example of this is involving psychiatry professionals with expertise in schizophrenia as integral components for chatbots targeting adults with schizophrenia.

\begin{figure}[ht!]
    \centering
    \includegraphics[width=0.9\linewidth]{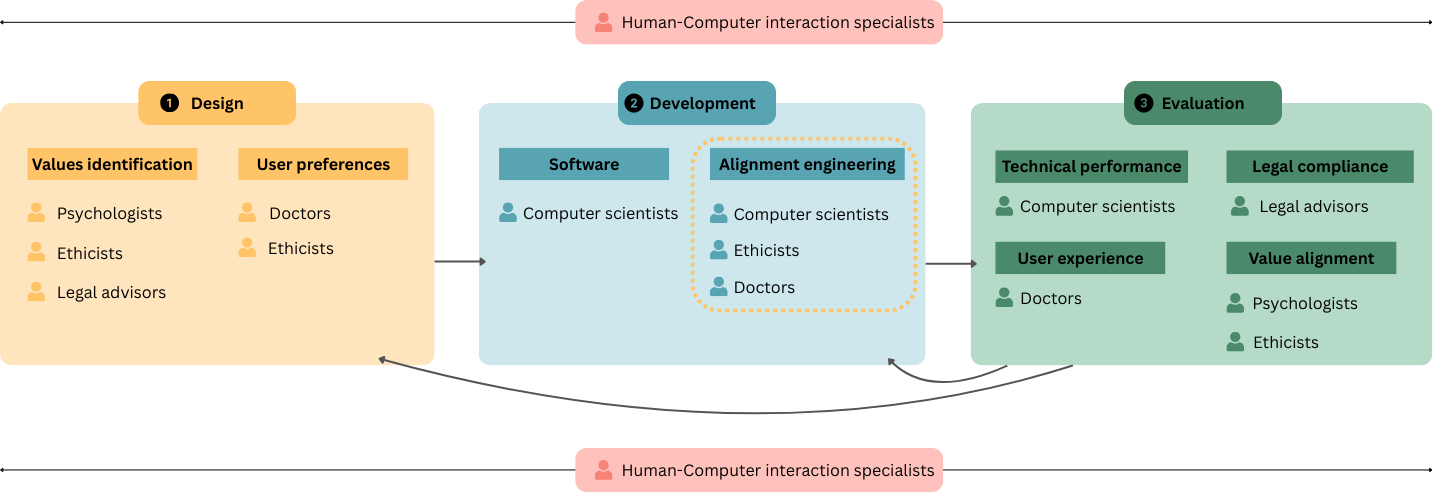}
    \caption{A deliberate interdisciplinary collaboration in a mental health chatbot.}
    \label{fig:case_study}
\end{figure}

To further elaborate on our position, Figure~\ref{fig:case_study} illustrates how phased contributions from different experts can potentially balance the cost-benefit trade-off. Technology experts guide the entire system implementation with HCI principles including human-in-the-loop methods and user-centric evaluations, translate values into technical constructs, develop the software, and assess the chatbot's technical performance. Mental healthcare professionals identify user needs, advise on implementation, and evaluate benefits, risks, and user satisfaction. Ethicists define ethical and social values, support implementation, and ensure value-alignment. Legal advisors specify regulatory requirements and ensure compliance. While we believe this interdisciplinary team is comprehensive and pragmatic, we acknowledge that the field is evolving and additional experts may be needed in the future.

\section{Recommendations}

Based on literature, we highlight a few guidelines and best-practices to balance the costs and benefits and hence enable interdisciplinary collaboration in mental health chatbots.\\
\noindent
\textbf{Interdisciplinary methodologies and frameworks:} The framework developed by \cite{hopkin2025toward} to categorize mental health AI systems could help clarify key characteristics, and thus relevant experts. A tangible interdisciplinarity evaluation measure (e.g., Rao–Stirling (\cite{leydesdorff2019interdisciplinarity})) and cross-sectoral methodologies could support effective interdisciplinary collaborations (\cite{boulos2021iterative}; \cite{molina2016role}). \\
\noindent
\textbf{EU AI Act:} Frameworks such as value sensitive design (\cite{friedman1996value}) and ethics-by-design (\cite{van2020designing}) can offer practical methods for interdisciplinary teams to integrate ethical and social values across the AI systems' lifecycle and anticipate risks in high-risk AI systems, as imposed by the AI Act.\\
\noindent
\textbf{Expert values and reflections:} Professionals should practice trust and mutual respect for differences, learn about methodologies from other disciplines, discuss any misconceptions and biases (\cite{pagliari2007design}); (\cite{boulos2021iterative}), and undertake training to familiarize themselves with different disciplines (\cite{moltrecht2022transdiagnostic}). \\

\section{Conclusion}
AI mental health chatbots require interdisciplinary collaboration for clinical benefit and value alignment. However, such collaborations face challenges and create a cost-benefit trade-off evident in today's chatbots, which have a narrow focus on experts and lifecycle phases. We argue for deliberately embedding technology, healthcare, ethics and law experts in a phased manner to guide the ethical and compliant implementation of mental health chatbots, with ethical design frameworks supporting this interdisciplinary collaboration across the lifecycle, value-alignment, and compliance with the high-risk AI Act requirements. While our position is theoretically sound, a critical future research dimension is the empirical analysis of the cost-benefit trade-off of involving different experts across the chatbot lifecycle, which will be necessary to develop concrete, practical recommendations for interdisciplinary collaboration.

\subsection*{Acknowledgments}
Funded by the European Union grants under the Marie Skłodowska-Curie Grant Agreements No: 101169473 (alignAI). Views and opinions expressed are, however, those of the author(s) only and do not necessarily reflect those of the European Union or the European Research Executive Agency (REA). Neither the European Union nor the European Research Executive Agency can be held responsible for them.

\bibliography{ais}

\end{document}